\documentclass{ws-ijmpcs}

\newcommand{\bea}{\begin{eqnarray}}
\newcommand{\ena}{\end{eqnarray}}

\newcommand{\be}{\begin{equation}}
\newcommand{\en}{\end{equation}}

\newcommand{\ed}{\end{document}}

\newcommand{\slp}{p\kern-5pt/}

\begin{document}

\markboth{Nurgul Habyl}
 {  }

%
\catchline{}{}{}{}{}
%

\title{Physical observables in the decay
$\Lambda_b \to \Lambda_c (\to \Lambda+\pi)+\tau^{-}+\bar \nu_{\tau}$ }

\author{N. Habyl}
\address{Al-Farabi Kazakh National University,
480012 Almaty, Kazakhstan\\
nuigui@mail.ru}
\author{Thomas Gutsche}
\address{
Institut f\"ur Theoretische Physik, Universit\"at T\"ubingen,
D-72076, T\"ubingen, Germany}
\author{Mikhail A. Ivanov}
\address{BLTP, 
Joint Institute for Nuclear Research, 141980 Dubna, Russia}
\author{J\"{u}rgen G. K\"{o}rner}
\address{Institut f\"{u}r Physik, 
Johannes Gutenberg-Universit\"{a}t, 
D-55099 Mainz, Germany}
\author{Valery E. Lyubovitskij}
\address{
Institut f\"ur Theoretische Physik, Universit\"at T\"ubingen,
T\"ubingen, Germany\\
Department of Physics, Tomsk State University,  
634050 Tomsk, Russia\\ 
Tomsk Polytechnic University, 
634050 Tomsk, Russia} 
\author{Pietro Santorelli}
\address{
Universit\`a di Napoli Federico II, 
80126 Napoli, Italy\\ 
INFN, Sezione di Napoli, 
80126 Napoli, Italy} 

\maketitle


\begin{abstract}
We analyze the tauonic semileptonic baryon decays 
$\Lambda^0_b \to \Lambda^+_c + \tau^{-} +\bar \nu_{\tau}$  with particular 
emphasis on the lepton helicity flip 
contributions which vanish for zero lepton masses. 
We calculate the total rate,
differential decay distributions, the longitudinal and transverse
polarization components of the $\Lambda^+_c$ and the $\tau^-$, 
and the lepton-side forward-backward asymmetries. 
%
%
We use the covariant confined quark model to provide numerical results on 
these observables.
\end{abstract}

\ccode{PACS numbers:12.39.Ki,13.30.Eg,14.20.Jn,14.20.Mr}

\section{Motivation}	

Recently there has been much discussion about tensions and discrepancies
of some of the experimental results on leptonic, 
semileptonic and rare decays involving (heavy) $\mu$ and $\tau$ leptons 
with the predictions of the Standard Model (SM).
Among these are the tauonic $B$ decays 
$B\to \tau \nu$, $B\to D\,\tau\bar\nu_\tau$ and
$B\to D^\ast\,\tau\bar\nu_\tau$ and the muonic decays 
$B \to K^{\ast}\mu^{+}\mu^{-}$
and 
$\mathrm{Br}[B \to K\mu^{+}\mu^{-}]/\mathrm{Br}[B \to Ke^{+}e^{-}]$. 
The situation has been nicely summarized in 
\cite{Soffer:2014kxa}\cdash\cite{Blake:2015tda}.
 
This observation has inspired a number of searches for new physics 
beyond the SM (BSM) in charged current interactions. Details can be
found in the recent literature on this subject (see, e.g. 
Refs.~\cite{Nierste:2008qe}\cdash\cite{Duraisamy:2014sna}).

Motivated by the discrepancy between theory and experiment in the meson 
sector we have analyzed in Ref.~\cite{Gutsche:2015mxa} the corresponding 
semileptonic baryon decays 
$\Lambda_b^0 \to \Lambda^+_c + \tau^{-} +\bar \nu_{\tau}$
within the SM with particular emphasis on the lepton helicity flip 
contributions which vanish for zero lepton masses. 
As in \cite{Korner:1987kd}\cdash\cite{Korner:1989qb} we have described the 
semileptonic decays using the helicity formalism which allows one to include 
lepton mass and polarization effects without much additional effort. 
We have calculated the total rate, differential decay distributions, 
the longitudinal and transverse polarization of the daughter baryon and 
lepton-side forward-backward asymmetries. 

Here, we give a brief sketch of the results obtained in 
Ref.~\cite{Gutsche:2015mxa} starting with the exact definition of
physical observables via helicity amplitudes squared.
Then we provide numerical results on these observables by using the 
covariant confined quark model. Also we replace an erroneous
factor of ``-3/2'' with the correct factor of ``-3/4'' in the definition
of the forward-backward asymmetry and provide correct numerical results
for this quantity.

\section{Helicity amplitudes and the polarization observables}

The matrix element of the process 
$\Lambda_b^0(p_1)\to \Lambda_c^+(p_2) + W^-_{\rm off-shell}(q)$ 
is expressed via the vector and axial vector current matrix elements 
which can be expanded in terms of a complete set of invariants

\begin{eqnarray}
M_\mu^V(\lambda_{1},\lambda_{2}) &=&
 \bar u_2(p_2,\lambda_{2})\bigg[F_1^V(q^2)\gamma_\mu-\frac{F_2^V(q^2)}{M_1}
  i\sigma_{\mu\nu}q^\nu+\frac{F_3^V(q^2)}{M_1}q_\mu\bigg]u_1(p_1,\lambda_{1}),
\nonumber\\
M_\mu^A(\lambda_{1},\lambda_{2}) &=& 
\bar u_2(p_2,\lambda_{2})\bigg[F_1^A(q^2)\gamma_\mu-\frac{F_2^A(q^2)}{M_1}
i\sigma_{\mu\nu}q^\nu+\frac{F_3^A(q^2)}{M_1}q_\mu\bigg]\gamma_{5}u_1(p_1,\lambda_{1})
\label{eq:inv2}
\end{eqnarray}
where $\sigma_{\mu\nu}=\frac i2(\gamma_\mu\gamma_\nu-\gamma_\nu\gamma_\mu)$ 
and  $q = p_1 - p_2$. The labels $\lambda_{i}=\pm \frac12$ denote the 
helicities of the two baryons.

\begin{figure}[pb]
\centerline{\includegraphics[width=6cm]{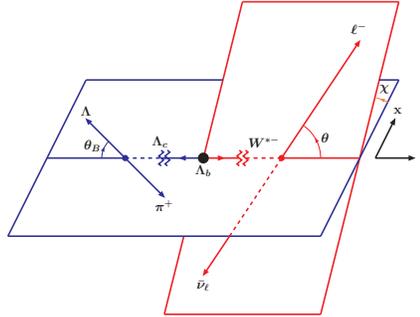}}
\vspace*{8pt}
\caption{Definition of the angles $\theta$,$\theta^*$ and $\chi$}
\label{fig:angles}
\end{figure}

It is easiest to calculate the helicity amplitudes in the rest frame of the 
parent baryon $B_{1}$ where we choose the $z$--axis to be along the 
$W^{-}_{\rm off-shell}$ (see Fig.~\ref{fig:angles}). 
They read \cite{Gutsche:2015mxa}

\begin{eqnarray}
H_{+\frac12 t}^{V/A} & = & \frac{\sqrt{Q_\pm}}{\sqrt{q^2}}
  \bigg( M_\mp F_1^{V/A}\pm \frac{q^2}{M_1} F_3^{V/A}\bigg), \nonumber\\
H_{+\frac12 +1}^{V/A} & = & \sqrt{2Q_\mp}
  \bigg(F_1^{V/A}\pm \frac{M_\pm}{M_1}F_2^{V/A}\bigg), \nonumber\\
 H_{+\frac12 0}^{V/A} & = & \frac{\sqrt{Q_\mp}}{\sqrt{q^2}}
  \bigg(M_\pm F_1^{V/A}\pm \frac{q^2}{M_1} F_2^{V/A}\bigg)\,. 
\label{eq:hel_inv}
\end{eqnarray}
where we make use of the abbreviations $M_\pm = M_1\pm M_2$ and 
$Q_\pm = M_\pm^2  - q^2$.

The physical observables can be expressed
in terms of helicity structure functions given in terms of  bilinear 
combinations of helicity amplitudes, see Table~I of 
Ref.~\cite{Gutsche:2015mxa}.  

One obtains the normalized differential rate expressed
in terms of the helicity structure functions 
\begin{eqnarray}
\frac{d\Gamma}{dq^2} &=& \Gamma_{0} \frac{(q^2-m_\ell^2)^2 
|\mathbf{p_2}|}{M_1^7q^2} \;
\Bigg\{  {\cal H}_U\, +\, {\cal H}_L\, 
+\,  \delta_\ell \, \Big[
 {\cal H}_U\, +\, {\cal H}_L\,+\, 3\,{\cal H}_S
\Big] \;\Bigg\} 
\nonumber\\
&\equiv& \Gamma_{0} \frac{(q^2-m_\ell^2)^2 |\mathbf{p_2}|}{M_1^7q^2} \; 
{\cal H}_{\rm tot}.
\label{eq:1-fold}
\end{eqnarray}
A forward-backward asymmetry is defined by
\begin{equation}
A_{FB}^{\ell}(q^2) =\frac{d\Gamma(F)-d\Gamma(B)}{d\Gamma(F)+d\Gamma(B)}
= -\frac34 \frac{H_P\,+4\,\delta_\ell\,H_{SL}}{H_{\rm tot}} .
\label{eq:FB}
\end{equation}
One defines a convexity parameter $C_F(q^2)$
according to
\begin{equation}
C_F(q^2)  = \frac34\,(1\,-\,2\delta_\ell)\, 
   \frac{H_U\,-\,2\,H_{L}}{H_{\rm tot}} .
\label{eq:CF}
\end{equation}

We obtain the $\cos\theta$ averaged polarization components of the daughter baryon $B_2$ 
($B_{2}=\Lambda^+_{c}$ in the present application).
One obtains 
\begin{eqnarray}
P^h_z(q^2) &=& \frac{\rho_{1/2\, 1/2}-\rho_{-1/2\, -1/2}}
                   {\rho_{1/2\, 1/2}+\rho_{-1/2\, -1/2}}
=\frac{ {\cal H}_{P} + {\cal H}_{L_P} 
+ \delta_{\ell}\,({\cal H}_{P} + {\cal H}_{L_P} + 3{\cal H}_{S_P})}{H_{\rm tot}}\,, 
\nonumber\\[1.5ex]
P^h_x(q^2) &=& \frac{2\,\mathrm{Re} \,\rho_{1/2\, -1/2}}
{\rho_{1/2\, 1/2}+\rho_{-1/2\, -1/2}}
=-\frac{3\pi}{4\sqrt{2}}\frac{{\cal H}_{LT} 
- 2\,\delta_{\ell}{\cal H}_{ST_P}}{{\cal H}_{\rm tot}}.
\label{eq:Pol_had}
\end{eqnarray}

We have calculated the $\cos\theta$ averaged polarization components of the lepton
with helicity flip contributions which can 
considerably change the magnitude of the polarization $|\vec {P}^{\ell}|$ 
and its orientation:
\begin{eqnarray}
P^\ell_z(q^2)  &=& 
-\frac{{\cal H}_U +{\cal H}_L-\delta_{\ell}\,( {\cal H}_U + {\cal H}_L 
+3{\cal H}_{S} )}{{\cal H}_{\rm tot}}\,, 
\nonumber \\[1.5ex]
P^\ell_x(q^2)  &=& -\frac{3\pi}{4\sqrt{2}}\sqrt{\delta_{\ell}}\,\,
\frac{{\cal H}_{P}-2\,{\cal H}_{SL}}{{\cal H}_{\rm tot}}.
\label{eq:Pol_lept}
\end{eqnarray}

The polarization of
the $\Lambda_{c}^{+}$ can be probed by analyzing the angular decay 
distribution of the subsequent decay of the $\Lambda_{c}^{+}$.
One  can exploit the cascade nature of the decay
$\Lambda_b^0 \to \Lambda_c^+ (\to \Lambda^0+\pi^+) 
+ W^-_{\rm off-shell}(\to \ell^- + \bar \nu_\ell)$
by writing down a joint angular decay distribution involving the polar angles 
$\theta,\, \theta_{B}$ and the azimuthal angles $\chi$ defined by the
decay products in their respective CM (center of mass) systems as shown in 
Fig.~\ref{fig:angles}.

\section{The transition form factors in the covariant confined quark model}

We shall use the covariant confined quark model
to describe the dynamics of the current--induced $\Lambda_b = (b[ud])$ to
$\Lambda_c = (c[ud])$ transition 
(see Refs.~\cite{Gutsche:2013pp,Gutsche:2013oea}).
The starting point of the model is an interaction Lagrangian
which describes the coupling of the $\Lambda_Q$-baryon to
the relevant interpolating three-quark current. One has
\begin{eqnarray}
{\cal L}^{\,\Lambda_Q}_{\rm int}(x) 
&=&g_{\Lambda_Q} \,\bar\Lambda_Q(x)\cdot J_{\Lambda_Q}(x) 
 + g_{\Lambda_Q} \,\bar J_{\Lambda_Q}(x)\cdot \Lambda_Q(x)\,,  
\label{eq:Lagr}\\[2ex]
J_{\Lambda_Q}(x) &=& \int\!\! dx_1 \!\! \int\!\! dx_2 \!\! \int\!\! dx_3 \, 
F_{\Lambda_Q}(x;x_1,x_2,x_3) \, J^{(\Lambda_Q)}_{3q}(x_1,x_2,x_3)\,,
\nonumber\\
J^{(\Lambda_Q)}_{3q}(x_1,x_2,x_3) &=& 
 \epsilon^{a_1a_2a_3} \, Q^{a_1}(x_1)\, u^{a_2}(x_2) 
\,C \, \gamma^5 \, d^{a_3}(x_3)\,,
\nonumber\\[2ex]
\bar J_{\Lambda_Q}(x) &=& 
\int\!\! dx_1 \!\! \int\!\! dx_2 \!\! \int\!\! dx_3 \, 
F_{\Lambda_Q}(x;x_1,x_2,x_3) \, \bar J^{(\Lambda_Q)}_{3q}(x_1,x_2,x_3)\,,
\nonumber\\
\bar J^{(\Lambda_Q)}_{3q}(x_1,x_2,x_3) &=& 
\epsilon^{a_1a_2a_3} \, \bar d^{a_3}(x_3)\, \gamma^5 \,C\, \bar u^{a_2}(x_2)  
\cdot \bar Q^{a_1}(x_1)\,. 
\nonumber
\end{eqnarray}

The form factors describing the $\Lambda_Q\to\Lambda_{Q'}$ transition
via the local weak quark current are calculated in terms of a two-loop 
Feynman diagram.
Due to the confinement mechanism of the model, the Feynman
diagrams do not contain branch points corresponding to on-shell quark 
production.

The results of our numerical two-loop calculation are well represented
by a double--pole parametrization
\begin{equation}
F(q^2)=\frac{F(0)}{1 - a s + b s^2}\,, \quad s=\frac{q^2}{M_1^2}
\end{equation}
with high accuracy: the relative error is less than 1\%.

In Table ~\ref{tab:tb_1} we list $q^2$ averaged helicity structure functions 
in units of  $10^{-15}$~GeV.
The numbers in Table~\ref{tab:tb_1} show that the results
of our dynamical calculation are very close to the HQET results
$\widetilde{\Gamma}_L=\widetilde{\Gamma}_S=-\widetilde{\Gamma}_{SL_{P}}$,
$\widetilde{\Gamma}_{ S_{P}}=\widetilde{\Gamma}_{L_{P}}
=-\widetilde{\Gamma}_{SL}$ and 
$\widetilde{\Gamma}_{ST_{P}}=\widetilde{\Gamma}_{LT}$. 
We do not display helicity flip results for the $e$ mode, 
because they are of order $10^{-6}-10^{-7}$ in the above units.

In Table ~\ref{tab:tb_2} we give the values of the integrated quantities. 
These can be obtained from the nonflip
and flip rates collected in Table ~\ref{tab:tb_1}.
In most of the shown cases, the mean values change considerably when going from 
the $e$ to the $\tau$ modes including even a sign change in~$<A_{FB}^{\ell}>$.
\begin{table}[ht]
\tbl{$q^2$ averaged helicity structure functions in units of $10^{-15}$ GeV.}
{\begin{tabular}{@{}ccccccc@{}} \toprule 
 & $ \Gamma_U $ & $  \Gamma_L $ & $ \Gamma_{LT} $ & $ \Gamma_P $ & 
$ \Gamma_{Lp} $ & $ \Gamma_{LTp} $ \\ \colrule
e & 12.4 & 19.6 & -7.73 & -7.61 & -18.5 & -3.50 \\
$ \tau $ & 3.29 & 2.90 & -2.06 & -1.73 & -2.46 & -0.66 \\ \colrule
 & $ \widetilde\Gamma_{U} $ & $ \widetilde\Gamma_{L} $ & 
$ \widetilde\Gamma_{S}$ & $ \widetilde\Gamma_{LT} $ & $ \widetilde\Gamma_{Sp} $ 
& $ \widetilde\Gamma_{SL} $ \\ 
$ \tau $ & 0.66 & 0.63 & 0.64 & -0.41 & -0.55 & 0.55 \\ \colrule
 & $\widetilde\Gamma_{P}$ & $\widetilde\Gamma_{Lp}$ &
     $ \widetilde\Gamma_{LTp} $ & $ \widetilde\Gamma_{STp} $ 
 & $ \widetilde\Gamma_{SLp} $ & \\ 
$\tau$ & -0.37 & -0.55 & -0.14 & -0.42 & -0.64 & \\
\botrule
\end{tabular} \label{tab:tb_1}}
\end{table}
\begin{table}[ph]
\tbl{The integrated quantities of physical observables}
{\begin{tabular}{@{}ccccccc@{}} \toprule 
 & $ <A_{FB}^{\ell}>$ & $ <C_F> $ & $ <P^h_z> $ & $ <P^h_x> $ & $ <P^\ell_z> $ & 
$ <P^\ell_x> $\\ \colrule
$e^-\bar\nu_e$       & 0.18     &  $-0.63$ & $-0.82$ & 0.40 & $-1.00$  & 0.00 \\
$\tau^-\bar\nu_\tau $ & $-0.038$ &  $-0.10$ & $-0.72$ & 0.22 & $-0.32$ & 0.55 \\
\botrule
\end{tabular} \label{tab:tb_2}}
\end{table}

\section*{Acknowledgments}

We are grateful to the organizers for the invitation to the conference
``Hadron Structure'15''.
This work was supported by the Tomsk State University Competitiveness
Improvement Program and the
Russian Federation program ``Nauka'' (Contract No. 0.1526.2015, 3854).
M.A.I. acknowledges the support from the Mainz Institute for 
Theoretical Physics (MITP). M.A.I. and J.G.K. thank the
Heisenberg-Landau Grant for support.



\end{document}